\documentclass[showpacs,amssymb,twocolumn,floatfix,pre,aps]{revtex4-1}

\usepackage{amsmath,amssymb,amsfonts,bm}
\usepackage{graphics,graphicx}

\DeclareMathOperator{\Tr}{Tr}

\begin{document}

\title{Reentrant disorder-disorder transitions in generalized
multicomponent Widom-Rowlinson models}

\author{Roman Kr\v{c}m\'ar}
\affiliation{Institute of Physics, Slovak Academy of Sciences, 
D\'ubravsk\'a cesta 9, SK-84511 Bratislava, Slovakia}
\author{Ladislav \v{S}amaj}
\affiliation{Institute of Physics, Slovak Academy of Sciences, 
D\'ubravsk\'a cesta 9, SK-84511 Bratislava, Slovakia}

\begin{abstract}
In the lattice version of the multicomponent Widom-Rowlinson (WR) model, 
each site can be either empty or singly occupied by one of $M$ different 
particles, all species having the same fugacity $z$.
The only nonzero interaction potential is a nearest-neighbor hard-core
exclusion between unlike particles.
For $M<M_0$ with some minimum $M_0$ dependent on the lattice structure, 
as $z$ increases from 0 to $\infty$ there is a direct transition from 
the disordered (gas) phase to a demixed (liquid) phase with one majority 
component at $z>z_d(M)$.
If $M\ge M_0$, there is an intermediate ordered 
``crystal phase'' (composed of two nonequivalent even and odd sublattices) 
for $z$ lying between $z_c(M)$ and $z_d(M)$ which is driven by entropy.
We generalize the multicomponent WR model by replacing the hard-core 
exclusion between unlike particles by more realistic large (but finite) 
repulsion.
The model is solved exactly on the Bethe lattice with an arbitrary 
coordination number. 
The numerical calculations, based on the corner transfer matrix 
renormalization group, are performed for the two-dimensional square lattice.
The results for $M=4$ indicate that the second-order phase transitions 
from the disordered gas to the demixed phase become of first order, for
an arbitrarily large finite repulsion.  
The results for $M\ge M_0$ show that, as the repulsion weakens, 
the region of crystal phase diminishes itself. 
For weak enough repulsions, the direct transition between the crystal and 
demixed phases changes into a separate pair of crystal-gas and gas-demixed 
transitions; this is an example of a disorder-disorder reentrant 
transition via an ordered crystal phase.
If the repulsion between unlike species is too weak, the crystal phase 
disappears from the phase diagram.
It is shown that the generalized WR model belongs to the Ising universality 
class.
\end{abstract}


\date{\today}

\maketitle

\section{Introduction} \label{Sec1}
The nearest-neighbors lattice gas analogy of Onsager's solution
of the two-dimensional Ising model enabled one to understand physical
implications of the spontaneous breaking of the particle-hole symmetry
on the existence and critical properties of high-density liquid and
low-density vapor phases \cite{Lee52}.

As concerns continuum fluids in thermal equilibrium, Widom and 
Rowlinson \cite{Widom70} introduced a simple model of identical 
particles (molecules) living in an infinite $\nu$-dimensional space of points 
${\bf r}\in V\to {\rm R}^{\nu}$.
There is a sphere of radius $R$ and volume $v_0$ around the center
of each molecule.
The potential energy $U$ associated with a given configuration 
${\bf r}_1,\ldots,{\bf r}_N$ of $N$ molecules is defined by
\begin{equation}
U({\bf r}_1,\ldots,{\bf r}_N) = \epsilon 
\left[ \frac{V({\bf r}_1,\ldots,{\bf r}_N)}{v_0} - N \right] , 
\end{equation}
where $V({\bf r}_1,\ldots,{\bf r}_N)$ denotes the volume covered by
the corresponding $N$ (in general penetrating) spheres and $\epsilon>0$
is some energy constant.
Due to the obvious inequalities 
$v_0\le V({\bf r}_1,\ldots,{\bf r}_N)\le N v_0$,
the potential energy is bounded as follows
\begin{equation}
-(N-1)\epsilon \le U({\bf r}_1,\ldots,{\bf r}_N) \le 0 . 
\end{equation}
The lower bound ensures a correct thermodynamics, the upper bound tells us
that the short-range forces among molecules are purely attractive.
The model is studied within the grand canonical ensemble characterized by
the dimensionless inverse temperature $\theta=\epsilon/(k_{\rm B}T)$ and
the particle fugacity $z$. 
The corresponding particle density is given by
$\rho(z,\theta) = v_0 \langle N \rangle/V$; $z$ is normalized so as to be 
asymptotically equal to $\rho$ in the ideal gas limit $\rho\to 0$.
   
The symmetry of the Widom-Rowlinson (WR) fluid, whose spontaneous breaking
is responsible for the existence of liquid and vapor phases, is hidden in
the original formulation.
It becomes transparent after mapping (in a thermodynamic sense) the WR model
onto a WR mixture of two kinds of molecules $\sigma\in\{ A,B\}$ interacting
in a pairwise manner $U(\{{\bf r}\}) = \sum_{i<j} 
u_{\sigma_i\sigma_j}(\vert {\bf r}_i-{\bf r}_j\vert)$, where the particles 
of the same species do not interact whereas the unlike species interact
with a hard-core repulsion at distances smaller than $2R$,
\begin{equation} \label{interaction}
u_{\sigma\sigma'}(r) = \left\{ 
\begin{array}{ll}
\infty & \mbox{if $\sigma\ne \sigma'$ and $r<2R$,} \cr
0 & \mbox{otherwise.}
\end{array} \right.
\end{equation}
Let us consider the case of equivalent species fugacities $z_A = z_B = z$.
At very low $z$, the system behaves like the ideal gas with just one pure
(mixed) phase with equivalent species densities $\rho_A = \rho_B$.
At very large $z$, since unlike molecules experience an infinitely strong
repulsion, the mixed phase suffers from packing effects which are 
substantially reduced in a demixed phase with a single majority component.
Consequently, the $A-B$ symmetry is broken and the WR mixture can exist
in two different homogeneous (i.e. translationally invariant) pure phases:
the $A$-rich phase with $\delta\rho\equiv \rho_A-\rho_B>0$ or the $B$-rich
phase with $\delta\rho<0$.
The two phases become equivalent ($\delta\rho=0$) at the ``demixing''
critical point $z_d$.
For dimensions $\nu\ge 2$, the proof of the existence of more than one pure
thermodynamic phase for sufficiently large $z$ was given by Ruelle
\cite{Ruelle71} using the Peierls contour method.
Ruelle's proof was generalized by Lebowitz and Lieb \cite{Lebowitz72} to 
the case when $u_{AB}(r)$ ($r<2R$) is large positive but not infinite.
Integral equation theories for the pair correlation functions of the WR 
mixture were developed in Refs. \cite{Yethiraj00,Brader07}.

The multicomponent generalization of the WR $A-B$ mixture consists in
considering molecules of $M$ different types $\sigma=1,\ldots,M$ 
with the same fugacity $z$.
The molecules interact in a pairwise manner, the only interaction is 
the hard-sphere repulsion between any two particles of unlike species 
like in (\ref{interaction}).
It was shown \cite{Lebowitz71,Runnels74} that in dimensions $\nu\ge 2$ 
the WR model with any finite number of components $M$ exhibits 
the demixing phase transition at some $z_d(M)$; in a pure demixed phase, 
the homogeneous density of just one of the components is dominant, 
say $\rho_1>\rho_2=\rho_3=\ldots=\rho_M$.
A hard (hyper-)cube version of the $M$-component WR model was studied in
the limit of infinite dimensionality $\nu\to\infty$ \cite{Sear96}; the
calculation of thermodynamic functions within the second virial coefficient
is exact in that limit.
It turns out that for $M\ge 31$ the transition from the mixed phase
at small values of $z$ to the demixed phase at large values of $z$ is
preempted by solidification at intermediate values of $z$,
$z_c(M) < z < z_d(M)$.
In the corresponding crystal phase, all species are equivalent
($\rho_1=\rho_2=\ldots=\rho_M=\rho/M$ with $\rho$ being the total density
of molecules), but the density $\rho\equiv \rho({\bf r})$ varies
periodically in space, i.e. the translational symmetry is broken.
The origin of this phenomenon is purely entropic: for large $M$ 
it pays the system to create a periodic structure of alternating dense 
and sparse regions, where the particles in the dense regions are less 
restricted by the hard-core repulsions coming from particles in 
the sparse regions.
 
In the lattice version of the multicomponent WR model 
\cite{Lebowitz71,Runnels74}, each lattice site $i$ can be either empty 
$\{ \sigma_i=0,z(0)=1\}$ or singly occupied by a particle of type 
$\sigma_i=1,2,\ldots,M$, all particles having the same fugacity 
$z(1)=z(2)=\ldots=z(M)\equiv z$.
The potential energy of a state configuration $\{ \sigma_i \}$ reads as
$U(\{\sigma_i\}) = \sum_{\langle i,j\rangle} u(\sigma_i,\sigma_j)$, where the
interaction potential between nearest-neighbor sites $\langle i,j\rangle$
is given by
\begin{equation} \label{intpot}
u(\sigma_i,\sigma_j) = \left\{ 
\begin{array}{ll}
\infty & \mbox{if $\sigma_i\ne \sigma_j$ and 
$\sigma_i\ne 0, \sigma_j\ne 0,$} \cr
0 & \mbox{otherwise.}
\end{array} \right.
\end{equation}
The number density of species $\sigma=1,\ldots,M$ at site $i$ will be denoted
by $\rho_i(\sigma)$, the total density of particles at site $i$ by
$\rho_i=\sum_{\sigma=1}^M \rho_i(\sigma)$.

The phase diagram of the lattice WR model is similar to that of 
the continuous WR model.
In dimensions $\nu\ge 2$ and for any finite number of components $M$,
the lattice WR model exhibits the demixing phase transition at some $z_d(M)$;
in a pure demixed phase, the site-independent density of just one of the
components is dominant, say $\rho(1)>\rho(2)=\ldots=\rho(M)$.
The effect of varying exclusion diameter between different species
was studied in \cite{Mazel14}.
When the number of components $M$ is equal to or larger than some minimum
$M_0$, an entropy-driven crystal phase exists for $z_c(M) < z < z_d(M)$;
the transition at $z_c(M)$ is always of second order.
In the crystal phase, the average total particle densities on the even and odd
sublattices are unequal, $\rho_e\ne \rho_o$, while the average densities of
the species $\sigma=1,2,\ldots,M$ are the same within a given sublattice,
i.e. $\rho_e(\sigma)=\rho_e/M$ and $\rho_o(\sigma)=\rho_o/M$.
The rigorous upper bound $M_0<27^6$ derived for the square lattice 
\cite{Runnels74} was surprisingly large.
The exact solution for the Bethe lattice of coordination $q$ gives 
$M_0 = [q/(q-2)]^2$ which would suggest more realistic $M_0\sim 4$ for $q=4$, 
while the Monte-Carlo (MC) simulations for the square lattice imply 
$M_0=7$ \cite{Lebowitz95}.
The extension of the hard-core exclusion to next-to-nearest-neighbors
leads to analogous phases whose numbers and characters depend on the specific
lattice geometry \cite{Georgii01}.

The infinite hard-core potential is a mathematical simplification of real 
inter-particle interactions.
In this paper, we generalize the multicomponent WR model by replacing 
the hard-core exclusion between unlike species by an arbitrary repulsion.
Namely, the infinity in the interaction potential (\ref{intpot}) between
unlike species on the nearest-neighbor sites is replaced by $U\ge 0$. 
The corresponding interaction Boltzmann factor
\begin{equation}
j = \exp(-\beta U)
\end{equation}
lies in the interval $[0,1]$.
The case $j=0$ $(U\to\infty)$ corresponds to the standard multicomponent 
WR model.
The opposite extreme case $j=1$ $(U=0)$ is equivalent to the non-interacting
$M$-component lattice gas of singly occupied lattice sites, with the trivial
grand partition function for a lattice of $N$ sites:
\begin{equation}
\Xi = (1+M z)^N . 
\end{equation}  
We shall concentrate on the neighborhood of the WR point $j=0$, i.e.
on large but finite repulsions, and study fundamental effects of nonzero 
$j$ on the phase diagram.

The model is solved exactly for the Bethe lattice with an arbitrary 
coordination number $q$. 
The numerical calculations, based on the corner transfer matrix 
renormalization group (CTMRG), are performed for the two-dimensional 
square lattice.
The results for $M=4$ indicate that the second-order phase transitions 
from the disordered gas to the demixed phase become of first order, for
an arbitrarily small positive $j$.  
The results for $M\ge M_0$ show that, as the repulsion weakens, 
the region of crystal phase diminishes itself. 
For weak enough repulsions, the direct transition between the crystal and 
demixed phases changes into a separate pair of crystal-gas and gas-demixed 
transitions; this is an example of a disorder-disorder reentrant 
transition via an ordered crystal phase.
If the repulsion between unlike species is too weak, the crystal phase 
disappears from the phase diagram.

A reentrant fluid-solid-fluid transition was observed in previous studies
of systems with soft cores and two repulsive interaction ranges, like
the two-scale ramp potential \cite{Hemmer70,Jagla99} or 
the square well--square shoulder model \cite{Kumar05,Lomba07}.
The computer simulations of the purely repulsive ramp potential
\cite{Franzese02,Skibinsky04} show a structural anomaly, namely the melting 
of the solid phase when the pressure is increased along an isotherm.
This phenomenon was detected also in Stillinger's Gaussian core model
\cite{Stillinger76,Lang00} and the antiferromagnetic Ising model with 
a nearest-neighbor interaction and a staggered mean field \cite{Hoye08},
and later on in models with one-scale interactions \cite{Saija09,Prestipino10}.
The structural anomaly is shared by real physical systems like water, silica
or phosphorus \cite{Brazhkin02,Shell02}.
Different definitions of the structural anomaly in fluids, which can lead to 
very different results, was discussed in \cite{Fomin14}.
In all mentioned papers, the fluid-solid and solid-fluid transitions are
discontinuous (of first order), with regions of the phase coexistence.
On the other hand, the disorder-crystal and crystal-disorder transitions of 
the generalized WR model are always continuous (of second order).

The paper is outlined as follows.
In Sec. II, we present the exact solution of the model on the Bethe
lattice with the coordination number $q$.
For the two-dimensional square lattice, the CTMRG technique 
is explained in Sec. III. A.
Sec. III. B brings the numerical results.
Conclusions are given in Sec. IV.

\section{Bethe lattice computation}
We consider the Bethe lattice of (locally equivalent) sites deep inside 
a tree, with coordination number $q$.  
The present calculation is based on the exact solution of 
an ``inverse problem'' for simply connected lattice structures 
\cite{Samaj89,Percus94}; for details in the $j=0$ case,
see Ref. \cite{Lebowitz95}. 

Every site $i$ of the Bethe lattice is an articulation point of
multiplicity $q$.
Site $i$ can be either empty, $\sigma=0$, or occupied by one of the
particles of $M$ different types, $\sigma=1,\ldots,M$.
The corresponding set of fugacities reads as $z_i(0)=1$ and 
$z_i(\sigma)=z$ for $\sigma=1,\ldots,M$.
For a given statistical model with two-site interactions between 
nearest-neighbor sites, we calculate the mean particle densities 
$\{ \rho_i(\sigma) \}$, constrained by
\begin{equation} \label{densityconstr}
\sum_{\sigma=0}^M \rho_i(\sigma) = 1 
\end{equation}
for each lattice site $i$.
The direct problem, find $\{ \rho_i(\sigma) \}$ for given 
$\{ z_i(\sigma) \}$, is nonlocal.
The inverse problem, find $\{ z_i(\sigma) \}$ for prescribed 
$\{ \rho_i(\sigma) \}$, is local in the sense that $z_i(\sigma)$ depends
on $\rho_i(\sigma)$ at the same site and on $\{ \rho_j(\sigma)\}$ 
at the nearest-neighbor sites $j=1,\ldots,q$.
According to \cite{Samaj89,Percus94}, the inverse solution for the Bethe
lattice can be constructed from local inverse problems for nearest-neighbor 
pairs of sites. 
\begin{itemize}
\item
Let $z_i^{\langle i,j\rangle}(\sigma)$ ($z_j^{\langle i,j\rangle}(\sigma)$) be the
set of fugacities at site $i$ $(j)$ which produce the given density
profiles $\rho_i(\sigma)$ and $\rho_j(\sigma)$ at nearest-neighbor sites 
$\langle i,j\rangle$.  
The fugacities at the original Bethe lattice are then expressible as 
\begin{equation} \label{origfugacity}
z_i(\sigma) = \left[ \frac{\rho_i(0)}{\rho_i(\sigma)} \right]^{q-1}
\prod_{j=1}^q z_i^{\langle i,j\rangle}(\sigma) .
\end{equation}
\item
Similarly, denoting by $\Xi^i$ the one-site and by $\Xi^{\langle i,j\rangle}$ 
the two-site grand partition function, the grand potential on the whole
lattice is given by
\begin{equation} \label{Omega}
\beta \Omega = - \sum_{\langle i,j\rangle} \ln \Xi^{\langle i,j\rangle} 
+ (q-1) \sum_i \ln \Xi^i .
\end{equation}
\end{itemize}
In the two-site problems, the expected symmetry breaking of the particle 
densities must be reflected via an analogous symmetry breaking in the
corresponding fugacities.
  
\subsection{Crystal phase}  
In the crystal phase, there are two alternating even and odd sublattices.
For the particle states $\sigma=1,\ldots,M$, we set $\rho_i(\sigma)=\rho_e$
for even sites and $\rho_i(\sigma)=\rho_o$ for odd sites.
In the two-site $\langle i,j\rangle$ problem, we have 
$z_i^{\langle i,j\rangle}(0) = z_j^{\langle i,j\rangle}(0) = 1$ and
$z_i^{\langle i,j\rangle}(\sigma) = z_1$, $z_j^{\langle i,j\rangle}(\sigma) = z_2$
for $\sigma=1,\ldots,M$.
The two-site grand partition function 
$\Xi^{\langle i,j\rangle} \equiv \Xi^{\langle 1,2\rangle}$ reads as
\begin{equation} \label{grand}
\Xi^{\langle 1,2\rangle} = 1 + M(z_1+z_2) + M z_1 z_2 + M(M-1) j z_1 z_2 ,
\end{equation} 
where the first term comes from two empty sites, the second one from one
empty site and the other occupied by an arbitrary particle, the third one 
from two particles in the same state and the fourth one from two particles 
in different states.
The corresponding one-site particle densities are given by
\begin{subequations}
\begin{eqnarray}
\rho_1 \Xi^{\langle 1,2\rangle} & = & z_1 + z_1 z_2 + (M-1) j z_1 z_2 , \\
\rho_2 \Xi^{\langle 1,2\rangle} & = & z_2 + z_1 z_2 + (M-1) j z_1 z_2 .
\end{eqnarray}
\end{subequations}
After some algebra, these equations determine $z_1$ and $z_2$ 
as functions of the particle densities as follows
\begin{subequations}
\begin{eqnarray}
z_1 & = & \frac{M(\rho_1+\rho_2)-1+(\rho_1-\rho_2)[1+j(M-1)]+\sqrt{D}}{
2(1-M\rho_1)[1+j(M-1)]} , \nonumber \\ & & \\
z_2 & = & \frac{M(\rho_1+\rho_2)-1+(\rho_2-\rho_1)[1+j(M-1)]+\sqrt{D}}{
2(1-M\rho_2)[1+j(M-1)]} , \nonumber \\ & & 
\end{eqnarray}
\end{subequations}
where the plus sign of the square root of the discriminant
\begin{eqnarray}
D & = & [1-(M-1)(1-j)(\rho_1+\rho_2)]^2 \nonumber \\ & &
+ 4\rho_1 \rho_2 (M-1)(1-j)[1+j(M-1)]
\end{eqnarray}
is fixed by the condition $z_{1,2}\to 0$ for $\rho_{1,2}\to 0$.
For $\Xi^{\langle 1,2\rangle}$, we get 
\begin{eqnarray} 
\Xi^{\langle 1,2\rangle} & = & \frac{M(M-1)(1-j)(\rho_1+\rho_2)}{
2(1-M\rho_1)(1-M\rho_2)[1+j(M-1)]} \nonumber \\
& & + \frac{2[1+j(M-1)] + M(\sqrt{D}-1)}{2(1-M\rho_1)(1-M\rho_2)[1+j(M-1)]} .
\end{eqnarray}
The original lattice fugacity in (\ref{origfugacity}) does not depend on site 
$i$ for all particle states $\sigma=1,\ldots,M$, $z_i(\sigma)=z$.
Using the density constraint (\ref{densityconstr}), we obtain 
one equation for each of two sublattices:
\begin{subequations}
\begin{eqnarray}
z & = & \left( \frac{1-M\rho_1}{\rho_1} \right)^{q-1} z_1^q , \label{1} \\
z & = & \left( \frac{1-M\rho_2}{\rho_2} \right)^{q-1} z_2^q . \label{2}
\end{eqnarray} 
\end{subequations}
The grand-potential per site $\beta\omega \equiv \beta\Omega/N$ is
determined from (\ref{Omega}) as 
\begin{equation}
\beta \omega = - \frac{q}{2} \ln\Xi^{\langle 1,2\rangle}
- \frac{q-1}{2} \ln \left[ (1-M\rho_1) (1-M\rho_2) \right] .
\end{equation}

Introducing the new variables $s=(\rho_1+\rho_2)/2$ and $t=\rho_1-\rho_2$,
Eqs. (\ref{1}) and (\ref{2}) can be written as
\begin{equation}
z = z(s,t) = z(s,-t) .
\end{equation}
They always have a trivial solution with $t=0$, which corresponds to
the disordered phase of equivalent even and odd sublattices, $\rho_1=\rho_2$.
A nontrivial solution $t\ne 0$ exists if $s\in [s_c^L,s_c^U]$, where
the lower and upper bounds are given by the equation 
$\partial z(s,t)/\partial t\Big\vert_{t=0} = 0$ \cite{Lebowitz95}:
\begin{subequations}
\begin{eqnarray}
s_c^L & = & \frac{1}{2M} \left( 1 - \sqrt{E} \right) , \\  
s_c^U & = & \frac{1}{2M} \left( 1 + \sqrt{E} \right)  
\end{eqnarray}
\end{subequations}
with 
\begin{equation}
E = 1 - \frac{4 M (q-1)}{(1-j) q^2 (M-1)} .
\end{equation}
The corresponding critical fugacities read
\begin{subequations}
\begin{eqnarray}
z_c^L & = & \frac{M^{q-1}}{[1+j(M-1)]^q} \frac{1-\sqrt{E}}{1+\sqrt{E}}
\left( \frac{(q-2)/q - \sqrt{E}}{1-\sqrt{E}} \right)^q , \nonumber \\  
& & \label{zcL} \\
z_c^U & = & \frac{M^{q-1}}{[1+j(M-1)]^q} \frac{1+\sqrt{E}}{1-\sqrt{E}}
\left(\frac{(q-2)/q + \sqrt{E}}{1+\sqrt{E}} \right)^q . \nonumber \\
& & \label{zcR} 
\end{eqnarray}
\end{subequations}
The value of the nontrivial $t$ changes continuously from 0 at $z\le z_c^L$
to some nonzero symmetry-broken value $\pm T$ in the interval $(z_c^L,z_c^U)$
and finally goes again continuously to 0 at $z_c^U$.
For $z>z_c^U$, the disordered phase with the trivial $t=0$ takes place. 

The crystal phase can exist only if the discriminant $E\ge 0$. 
This means that, for the fixed coordination number $q$ and the
Boltzmann factor $j<(q-2)^2/q^2$, the number of components $M$ 
must be equal or larger than some minimum $M_0$, $M\ge M_0$, given by
\begin{equation} \label{defM0}
M_0 = \frac{(1-j) q^2}{(q-2)^2-j q^2} .
\end{equation}
Note that $M_0\to\infty$ just at $j=(q-2)^2/q^2$.
Equivalently, for the fixed coordination number $q$ and the number of 
components $M>(q-2)^2/q^2$, the crystal phase exists only if the Boltzmann 
factor $j\le j_{\max}$, where the maximum value $j_{\max}$ is given by
\begin{equation} \label{jmax}
j_{\max} = 1 - \frac{4 M (q-1)}{q^2(M-1)} .
\end{equation}
In the limit of $M\to\infty$, we have the asymptotic expansion
\begin{equation} \label{jmaxasym}
j_{\max} = \frac{(q-2)^2}{q^2} - \frac{4(q-1)}{q^2} \frac{1}{M} 
+ O\left( \frac{1}{M^2} \right) .
\end{equation}

\subsection{Demixed phase}  
In the demixed phase, all sites are equivalent but one of the components,
say $\sigma=1$, is dominant.
This means that the density of 1-particles $\rho_i(1)=\rho(1)$ is larger than 
$\rho_i(\sigma)=\rho(2)$ for all remaining particle states $\sigma=2,\ldots,M$. 
In the two-site $\langle i,j\rangle$ problem, we set 
$z_i^{\langle i,j\rangle}(0) = z_j^{\langle i,j\rangle}(0) = 1$, 
$z_i^{\langle i,j\rangle}(1) = z_j^{\langle i,j\rangle}(1) = z(1)$ for particles of
type 1 and $z_i^{\langle i,j\rangle}(\sigma) = z_j^{\langle i,j\rangle}(\sigma) = z(2)$
for all particles of type $\sigma=2,\ldots,M$.
The two-site grand partition function is given by
\begin{eqnarray}
\Xi^{\langle 1,2\rangle} & = & 1 + 2 z(1) + z(1)^2 + 2(M-1) z(2) \nonumber \\
& & + (M-1) z^2(2) + 2 (M-1) j z(1) z(2) \nonumber \\ & &  
+ (M-1)(M-2) j z^2(2) .
\end{eqnarray}
The corresponding particle densities are given by
\begin{subequations}
\begin{eqnarray}
\rho(1) \Xi^{\langle 1,2\rangle} & = & z(1) + z^2(1) + (M-1) j z(1) z(2) , \\
\rho(2) \Xi^{\langle 1,2\rangle} & = & z(2) + z^2(2) + j z(1) z(2) 
+ (M-2) j z^2(2) . \nonumber \\ & &
\end{eqnarray}
\end{subequations}
Considering in (\ref{origfugacity}) $z_i(\sigma)=z$ for all sites $i$ and
particle states $\sigma=1,2,\ldots,M$, we get
\begin{subequations}
\begin{eqnarray}
z & = & \left( \frac{1-\rho(1)-(M-1)\rho(2)}{\rho(1)} \right)^{q-1} z(1)^q , 
\label{l1} \\
z & = & \left( \frac{1-\rho(1)-(M-1)\rho(2)}{\rho(2)} \right)^{q-1} z(2)^q . 
\label{l2}
\end{eqnarray} 
\end{subequations}
The grand-potential per site is obtained in the form
\begin{equation}
\beta \omega = - \frac{q}{2} \ln\Xi^{\langle 1,2\rangle}
- (q-1) \ln \left[ 1 - \rho(1) - (M-1)\rho(2) \right] .
\end{equation}
The above set of nonlinear equations can be solved only numerically.
The trivial disordered solution $\rho(1)=\rho(2)$ always exists.
If more real solutions exist, the one with the minimal $\beta\omega$
dominates.

Let $M_t\le M_0$ denote the ``transition'' number of components, such that
the direct transition from the disordered to the demixed phase is of 
second order for $M\le M_t$ and of first order for $M>M_t$.
For the Bethe lattice of the coordination number $q$, we have 
the trivial value $M_t=2$ independent of $j$. 

\begin{figure}
\includegraphics[width=0.4\textwidth,clip]{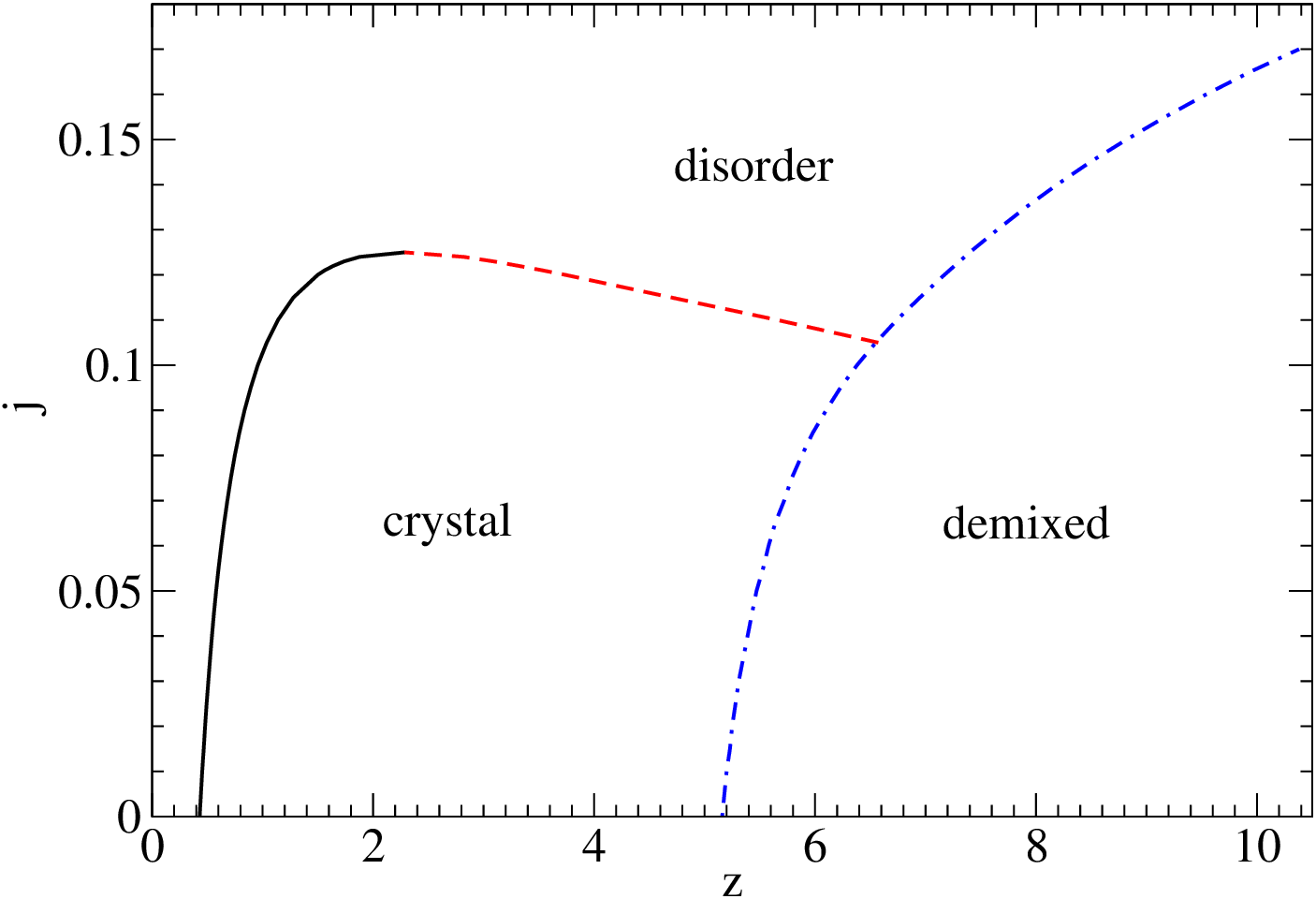}
\caption{\footnotesize (Color online) The phase diagram in the $(z,j)$ plane
for the generalized WR model on the Bethe lattice with the coordination 
number $q=4$ and the number of components $M=7$.}
\label{fig:Bethe}
\end{figure}

In Fig. \ref{fig:Bethe}, we present the phase diagram in the $(z,j)$ plane
for our generalized WR model on the Bethe lattice with the coordination 
number $q=4$ and the number of components $M=7$. 
Note that from Eq. (\ref{defM0}) we have $M_0=4$ for $j=0$ and $M_0=7$ 
for $j=1/8$, i.e. there is no crystal phase for $M=7$ components if $j>1/8$.
The second-order transition line from the disorder to crystal phases,
given by formula (\ref{zcL}), is depicted by the solid curve.
The second-order transition line from the crystal to disorder phases,
given by formula (\ref{zcR}), is depicted by the dashed curve; 
along this line the system exhibits the reentrant phenomenon.
The transitions from the crystal or disordered phases to the demixed phase
take place along the dash-dotted curve.
It is seen that for small $j\lesssim 0.105$, the successive order in which 
the phase transitions take place is basically the same as in the standard 
WR model with $j=0$.
For weak enough repulsions $j\in [0.105,0.125)$, the direct transition 
between the crystal and demixed phases changes into a separate pair 
of crystal-gas and gas-demixed transitions; this is an example of 
a disorder-disorder reentrant transition via an ordered crystal phase.
If the repulsion between unlike species is too weak $j>1/8$, 
the crystal phase disappears from the phase diagram.
     
\section{CTMRG method}
\subsection{Technique}
On the square lattice, the free energy $\mathcal{Z}$ can be decomposed into 
four corner transfer matrices $\mathcal{C}$, each representing the 
Boltzmann weight of a quadrant of the lattice system 
$\mathcal{Z} = \Tr\mathcal{C}^4$  \cite{Baxter}. 
The fourth power of the corner transfer matrix can be then interpreted 
as the density matrix $\rho = \mathcal{C}^4$. 
The concept of the renormalization can be applied to this density matrix 
\cite{white1,white2,schollwock}. 
The CTMRG method combines the corner transfer matrix representation of 
the density  matrix, the density matrix renormalization and an accurate 
approximation of the free energy for a large scale system in terms of 
an iterative numerical calculation \cite{tomo1,tomo2,tomo3}. 
During the process, the space of states is truncated.
The dimension of the truncated space is denoted by $D$; the larger value
of $D$ is used, the more accurate results are obtained.

We consider a square of lattice sites with dimension $L\times L$.
There is a central row and $N$ additional rows on both sides,
so $L = 2N+1$ where $N$ counts for the number of iterations of 
the transfer matrix. 
When the density matrix $\rho^{(N)}$ is obtained for a sufficiently 
large system, we calculate the expectation values of microscopic variables 
at the center of the system, which represents bulk thermodynamics deep 
inside the system. 
For example, the spontaneous particle density for our crystal phase 
is obtained as follows
\begin{equation} \label{mc}
m^{(N)}_c = \frac{1}{\Tr\rho^{(N)}} \Tr\left\{ [2n(i,j)-1] (-1)^{i+j}\rho^{(N)}
\right\} ,
\end{equation}
where $\{i,j\}$ are the coordinates of the central point and the function 
$n(i,j)=0$ if the central site is empty and $n(i,j)=1$ otherwise.
This expression reproduces correctly $m_c$ as the sublattices $A$ and $B$
difference $m_A-m_B$.
As the order parameter for the demixed phase, we consider
\begin{equation} \label{md}
m_d = \left\vert m_1 {\rm e}^{-{\rm i}\phi_1} + \cdots +
m_M {\rm e}^{-{\rm i}\phi_M} \right\vert ,
\end{equation}
where $m_{\sigma}$ is the occupation of the particle state $\sigma=1,2,\ldots,M$
and the angle $\phi_{\sigma} = 2\pi(\sigma-1)/M$.
The quantity $m_d$ vanishes in the disordered phase with $m_1=m_2=\cdots=m_M$
and attains a positive value if one of the components is dominant.
In the limit $z\to\infty$, $m_d=1$ for any value of $M\ge 2$ and $j\in [0,1]$.

In order to detect the position of the phase transitions, special quantities 
exhibiting singular behavior near the phase transition, like the specific 
heat, are usually used. 
Here, we use the von Neumann entropy, defined as
\begin{equation}
S_N = -\Tr\rho\ln\rho .
\end{equation}
Close to a second-order critical point, the von Neumann entropy can be 
expressed as \cite{calabrese,ercolessi}
\begin{equation} \label{eq:cko}
S_N \sim \frac{c}{6} \ln{\xi},
\end{equation}
where $c$ is the central charge and $\xi$ is the correlation length
of the particle system. 
Consequently, the von Neumann entropy has a logarithmic divergence
at the critical point. 
At a first-order transition point, it exhibits a discontinuity. 

As concerns critical exponents, we shall calculate the plot of 
the spontaneous density around the critical point to obtain 
the exponent $\beta$.
Moreover, in two dimensions and at the critical point, we shall study 
the finite-size scaling of the particle density 
\begin{equation}
m(L) \propto L^{-\eta/2} , \qquad \mbox{large sample size $L$,}
\end{equation}
to deduce the exponent $\eta$ \cite{tomo2}. 
Using, e.g., the numerical logarithmic derivative
\begin{equation} \label{eq:etaeff}
\eta_{eff}(L) = -2\frac{\ln[m(L+2)]-\ln[m(L)]}{\ln(L+2)-\ln(L)} ,
\end{equation}
we can extract the critical exponent $\eta$ as the limit
$\eta = \lim_{L\to\infty} \eta_{eff}(L)$. 

\subsection{Numerical results}
\begin{figure}
\includegraphics[width=0.4\textwidth,clip]{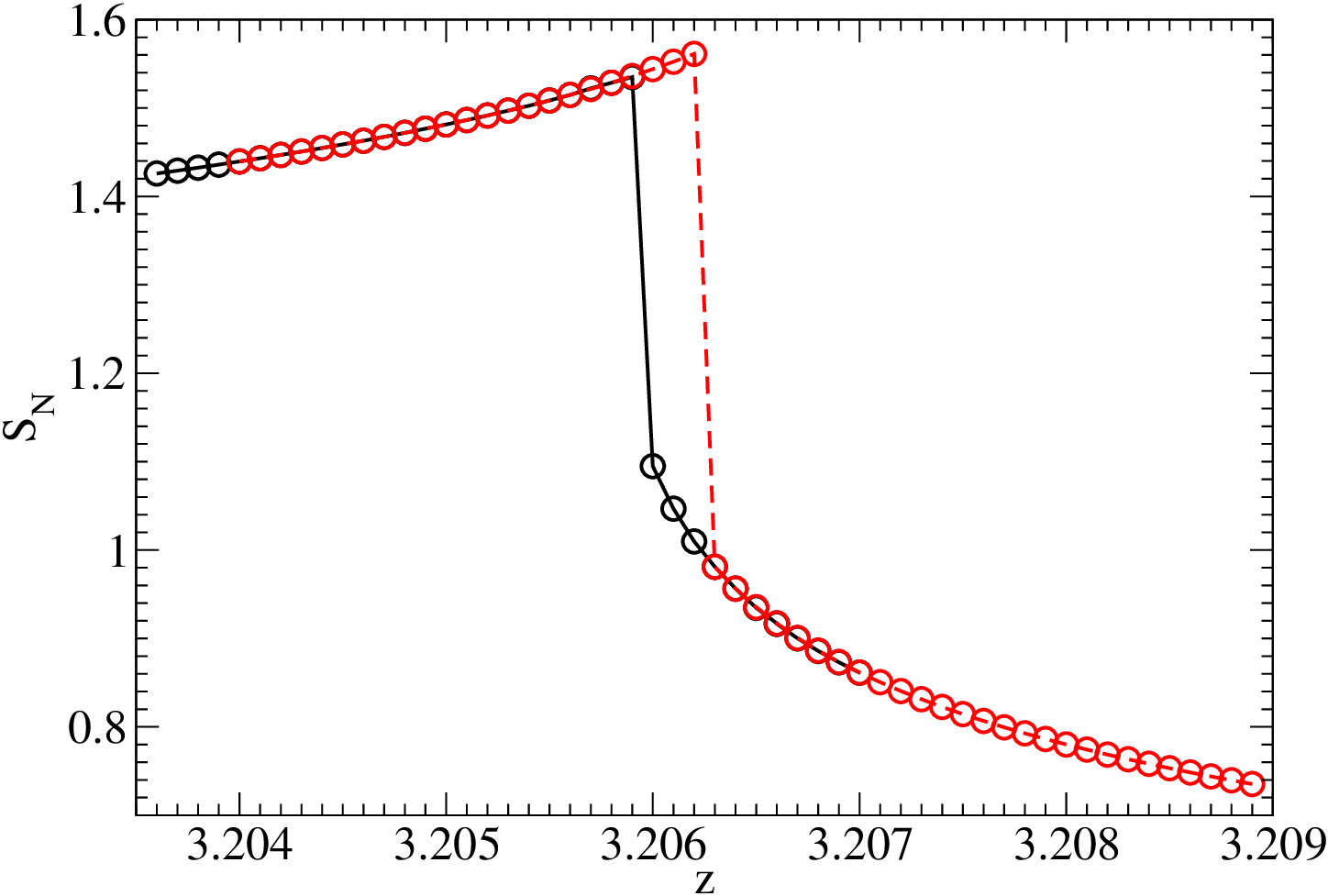}
\caption{\footnotesize (Color online) 
The von Neumann entropy of the disorder-demixed phase transition for 
the generalized WR model with $M=4$ components and a large finite 
repulsion $j=0.001$. 
The solid and dashed curves correspond to fixed and free boundary conditions, 
respectively.}
\label{fig:entropy}
\end{figure}

\begin{figure}
\includegraphics[width=0.4\textwidth,clip]{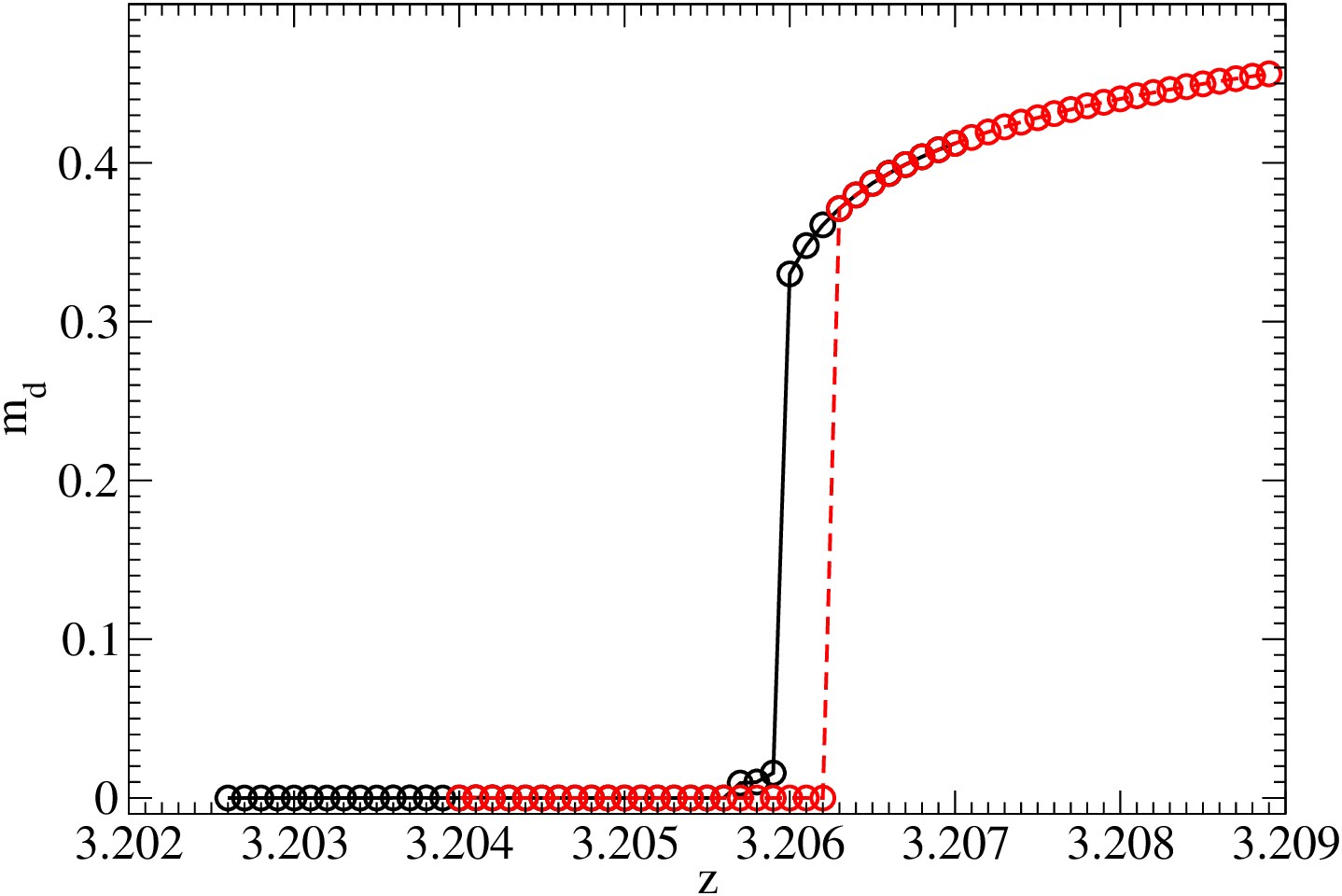}
\caption{\footnotesize (Color online) The spontaneous particle density of 
the disorder-demixed transition for $M=4$ and $j=0.001$.}
\label{fig:magnetization}
\end{figure}

The MC simulations for $j=0$ in Ref. \cite{Lebowitz95} indicate that
at the direct disorder-demixed phase transition $z_d$, there is no jump
in the density for $M\le 4$ (so the transition is of second-order) while
there is a jump in density for $M\ge 5$.
In other words, $M_t=4$ for the standard WR model on the square lattice.
For $M=4$ particle components, an arbitrarily large (but finite) repulsion
between different species causes that the disorder-demixed transition
becomes of first order.
This fact is documented in Fig. \ref{fig:entropy} on the plot of 
the von Neumann entropy versus the fugacity for the generalized WR model 
with $M=4$ components and a large finite repulsion $j=0.001$.
The solid (black) curve connects data evaluated with boundary conditions 
fixed at the particle state dominant in the demixed phase while the dashed 
(red) curve connects data for free boundary conditions. 
We see that the two curves calculated with distinct boundary conditions
produce a hysteresis inside which the disorder and demixed phase coexists,
as is usual in the case first-order phase transitions \cite{Graf98}.
The corresponding plot of the order parameter $m_d$ (\ref{md}) is
presented in Fig. \ref{fig:magnetization}.
For $M=3$ components, the second-order phase transition takes place
also for small $j>0$.

\begin{figure}
\includegraphics[width=0.4\textwidth,clip]{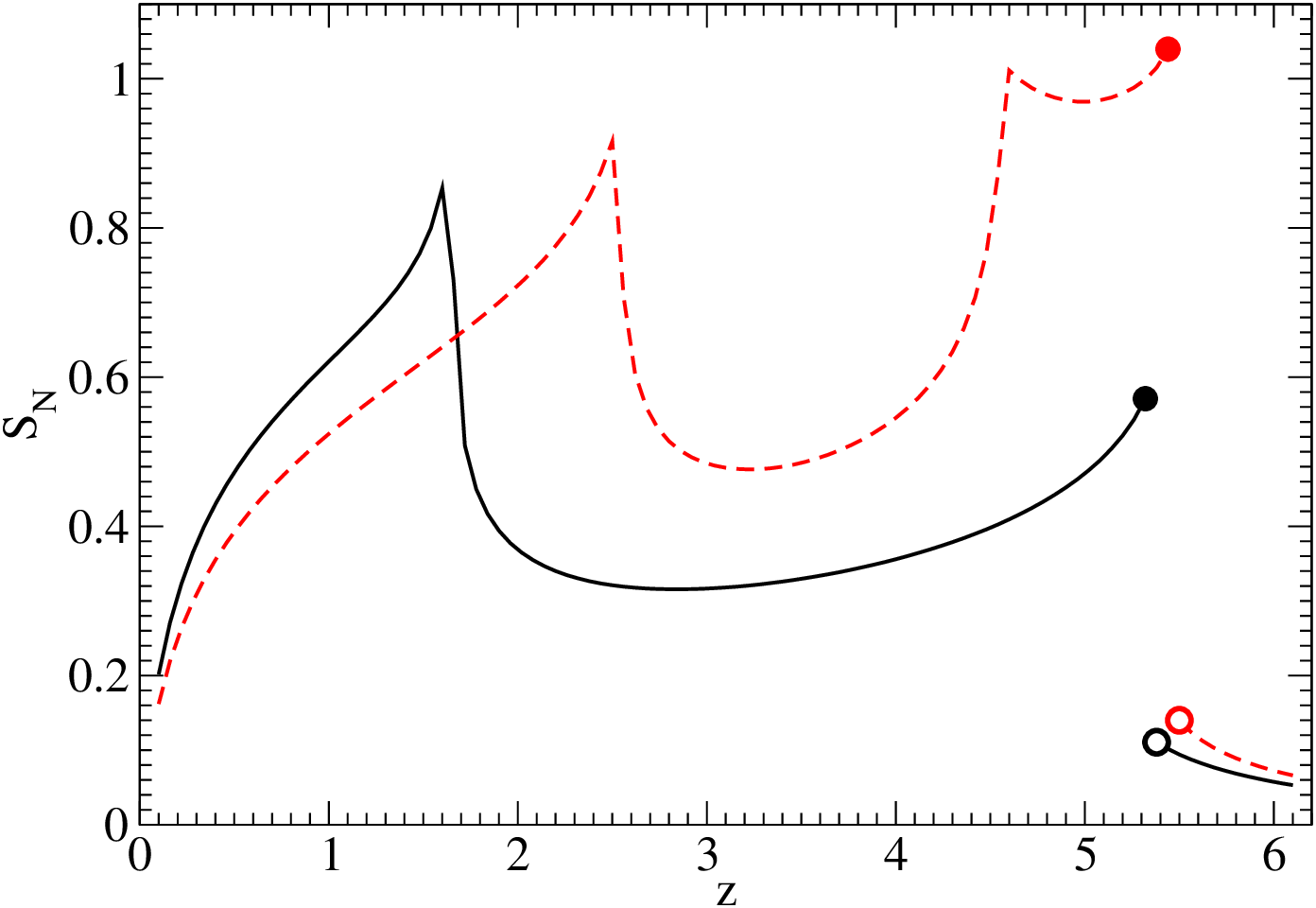}
\caption{\footnotesize (Color online) 
The von Neumann entropy $S_N$ versus the fugacity $z$ for the WR model with 
$M = 7$ particle components. 
The solid curve corresponds to an infinite $j = 0$ repulsion between unlike
species, the dashed curve corresponds to a finite repulsion $j=0.018$.
Dimension of truncated space of states is $D=50$.}
\label{fig:tetra1}
\end{figure}

\begin{figure}
\includegraphics[width=0.4\textwidth,clip]{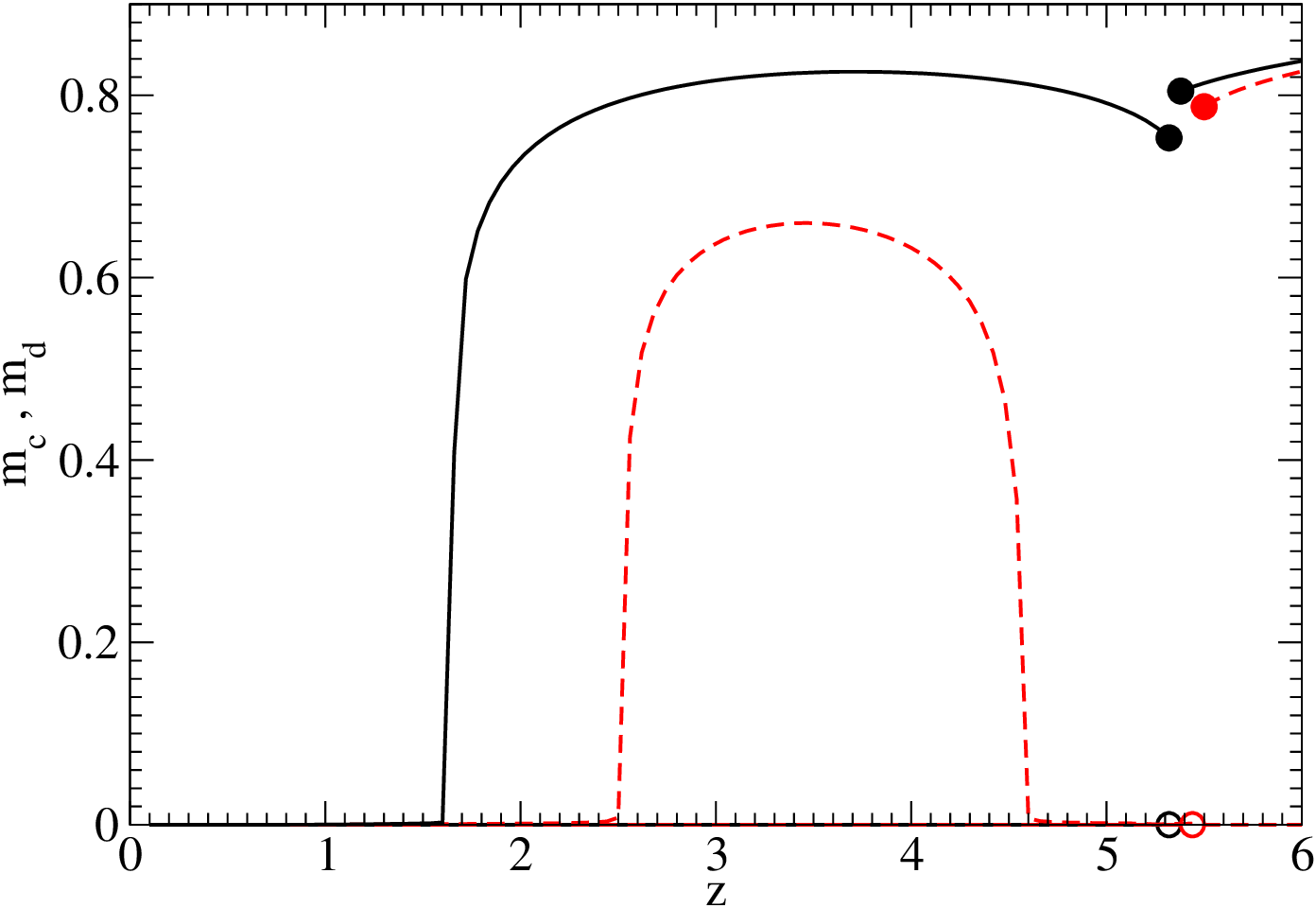}
\caption{\footnotesize (Color online) 
The crystal order parameter $m_c$ (left fragment) and the demixed order 
parameter $m_d$ (right fragment) for the Widom-Rowlinson model with $M = 7$ 
components.
The solid and dashed curves correspond to $j = 0$ and $j=0.018$, respectively.
$D = 50$.}
\label{fig:tetra2}
\end{figure}

The MC simulations for $j=0$ in Ref. \cite{Lebowitz95} indicate that
the standard WR model on the square lattice exhibits the crystal phase at
and beyond $M_0=7$ particle components.
In Fig. \ref{fig:tetra1}, we plot by solid curve the von Neumann entropy 
$S_N$ versus the fugacity $z$ for that standard $j=0$ WR model with 
$M = 7$ components.
There are two singularities.
The logarithmic one on the left indicates the continuous (second-order) 
disorder-crystal phase transition, the right one indicates the discontinuous 
(first-order) crystal-demixed phase transition. 
For the generalized WR model with $j=0.018$ (dashed curve), there are 
three singularities.
The one in the middle corresponds to the reentrant transition from the crystal 
to gas phase.
The corresponding plots of the order parameters $m_c$ (\ref{mc}) and
$m_d$ (\ref{md}) are presented in Fig. \ref{fig:tetra2}.

\begin{figure}
\includegraphics[width=0.4\textwidth,clip]{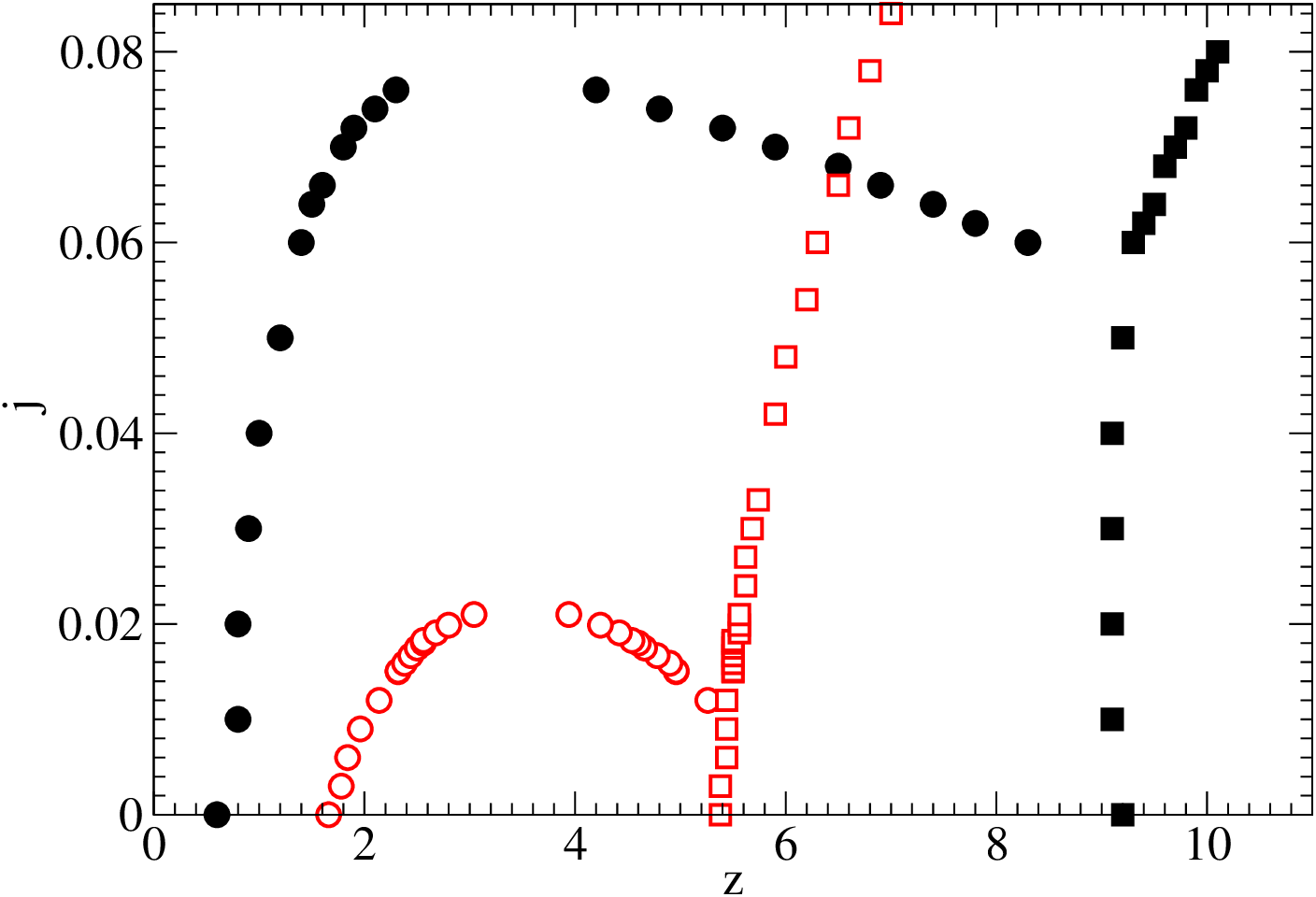}
\caption{\footnotesize (Color online) 
The critical lines in the $(z,j)$ plane for the generalized WR model
with $M = 7$ (open symbols) and $M = 10$ (full symbols) components. 
$D = 30$.}
\label{fig:tetra4}
\end{figure}

In Fig. \ref{fig:tetra4}, the phase diagrams in the $(z,j)$ plane
are presented for the generalized WR model with $M = 7$ (open symbols) 
and $M = 10$ (full symbols) components. 
The disorder-crystal and crystal-disorder phase transitions are
represented by circles, the crystal-demixed and disorder-demixed phase
transitions are denoted by squares.
The phase diagrams resemble the one for the Bethe lattice in
Fig. \ref{fig:Bethe}.

\begin{figure}
\includegraphics[width=0.4\textwidth,clip]{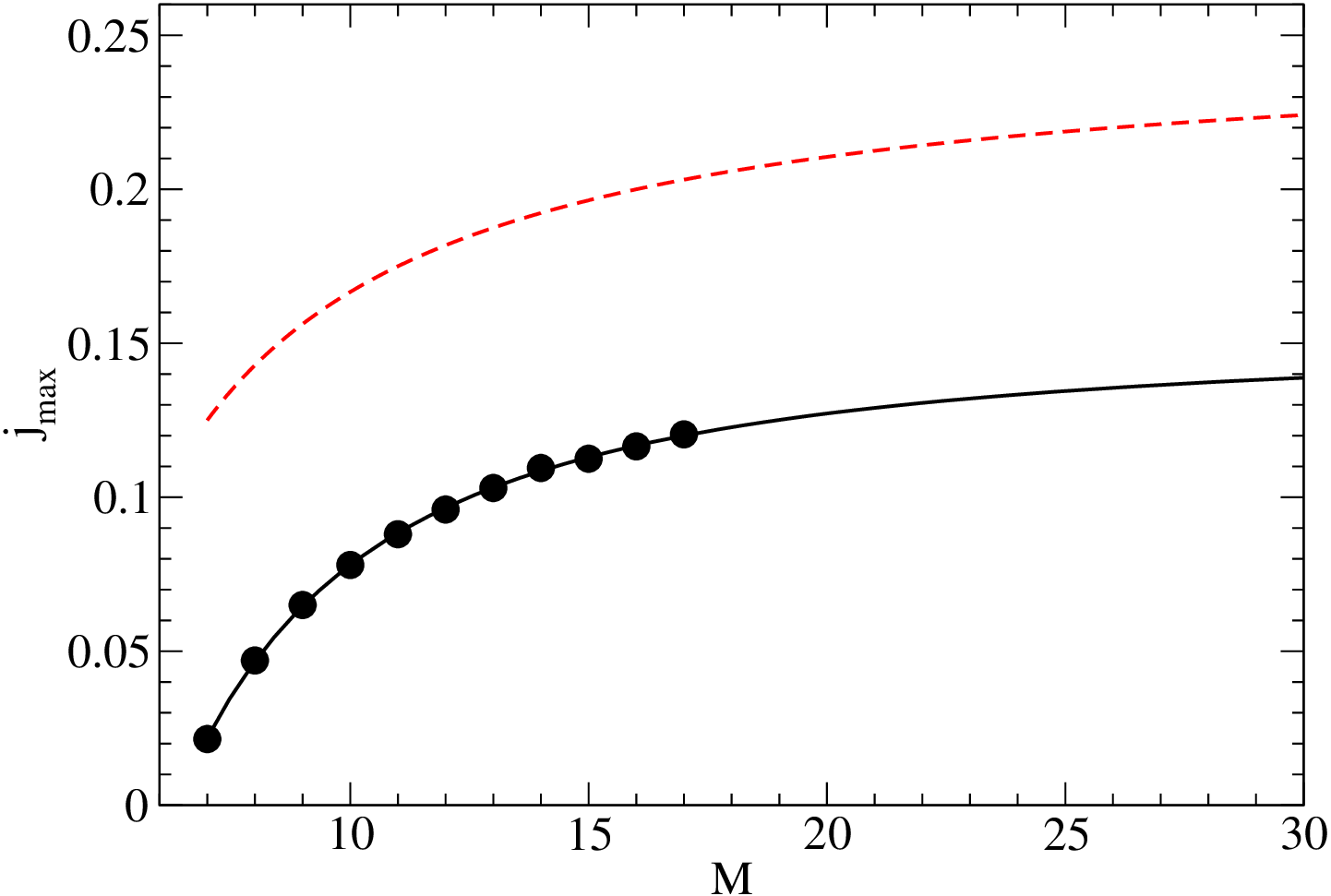}
\caption{\footnotesize (Color online) 
The maximal repulsion Boltzmann factor $j$ for which the crystal phase exists 
in the generalized WR model on the square lattice with $M$ components. 
The dashed curve corresponds to the exact Bethe result (\ref{jmax}) for
the coordination number $q=4$.
The solid curve is the (inverse) polynomial fit of the numerical CTMRG data
(solid circles) which converges to $j_{\max}=0.15$ as $M\to \infty$. 
$D = 30$.}
\label{fig:octo2}
\end{figure}

For the generalized $M$-component WR model on the Bethe lattice with 
the coordination number $q$ in Sec. II, we have defined the maximal 
repulsion Boltzmann factor $j_{\max}$, for which the crystal phase exists.  
For $q=4$, the exact formula (\ref{jmax}) implies
\begin{equation}
j_{\max} = \frac{1}{4} - \frac{3}{4} \frac{1}{M-1}
\mathop{\sim}_{M\to\infty} \frac{1}{4} - \frac{3}{4} \frac{1}{M} .
\end{equation}
The corresponding dependence of $j_{\max}$ on the number of components $M$
for the Bethe lattice with $q=4$ is represented in Fig. \ref{fig:octo2} 
by dashed curve.
The solid curve is the (inverse) polynomial fit of the numerical CTMRG data
for the square lattice (solid circles).
In the asymptotic limit $M\to \infty$, the fit implies that
\begin{equation}
j_{\max} \mathop{\sim}_{M\to\infty} 0.15 - \frac{2.9}{M^{1.6}} .
\end{equation}

\begin{figure}
\includegraphics[width=0.4\textwidth,clip]{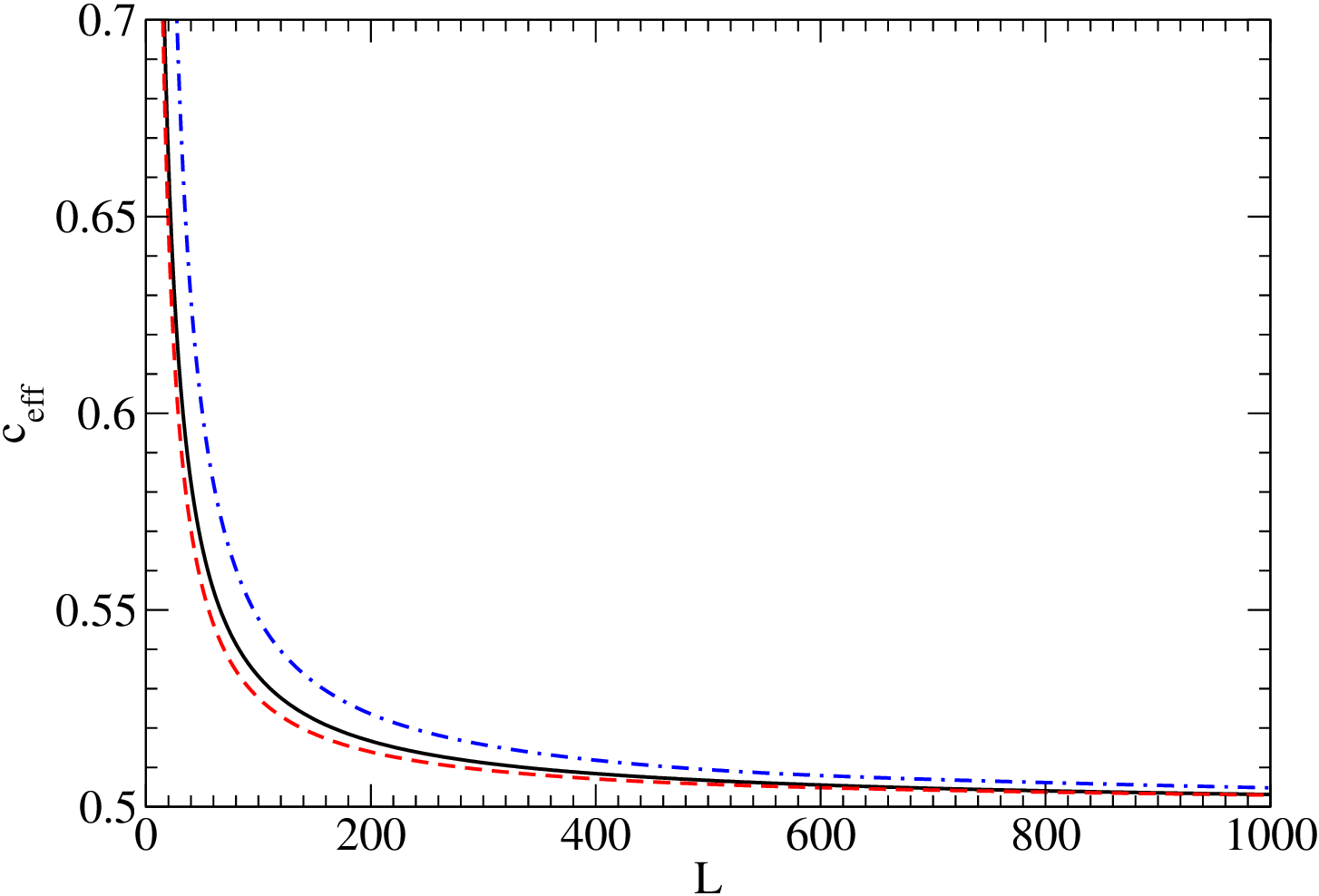}
\caption{\footnotesize (Color online) 
The WR model on the square lattice with $M=7$ components.
The dependence of $c_{eff}$ (\ref{ceff}) on the system size $L$.
The solid curve stands for the standard $j=0$ WR model at $z_c=1.66196$,
the dashed curve corresponds to $(j=0.018, z_c^L=2.58545)$ and finally
the dash-dotted curve corresponds to $(j=0.018, z_c^U=4.47455)$.
As $L\to\infty$, all three curves converge to the central charge $c=1/2$ 
of the Ising universality class.
$D=300$.}
\label{fig:octo4}
\end{figure}

At the critical point, the correlation length of the particle system $\xi$
diverges in the bulk and $\xi\propto L$ for a finite system of characteristic 
length $L$.
According to Eq. (\ref{eq:cko}), defining
\begin{equation} \label{ceff}
c_{eff}(L) = 6 \frac{\partial S_N}{\partial (\ln L)} ,
\end{equation}
the central charge $c$ of the critical model is obtained as the limit
$c = \lim_{L\to\infty} c_{eff}(L)$. 
In what follows, we shall restrict ourselves to the WR model on the square
lattice with $M=7$ components.
For $j=0$, there is the only critical point $z_c=1.66196$ at which
the disorder-crystal phase transition takes place.
For $j=0.018$, there is the disorder-crystal phase transition at $z_c^L=2.58545$
and the reentrant crystal-disorder phase transition at $z_c^U=4.47455$.
In Fig. \ref{fig:octo4}, we plot the dependence of $c_{eff}$ on 
the system size $L$ for the above three critical points:
the solid curve stands for $(j=0, z_c=1.66196)$, the dashed curve corresponds 
to $(j=0.018, z_c^L=2.58545)$ and the dash-dotted curve corresponds to 
$(j=0.018, z_c^U=4.47455)$.
It is seen that as $L$ goes to infinity, all three curves converge to 
the value $c=1/2$ which is the central charge of the Ising universality class.
Note that due to relatively large values of $L$ we have to increase 
the dimension of the truncated space to $D=300$.

\begin{figure}
\includegraphics[width=0.4\textwidth,clip]{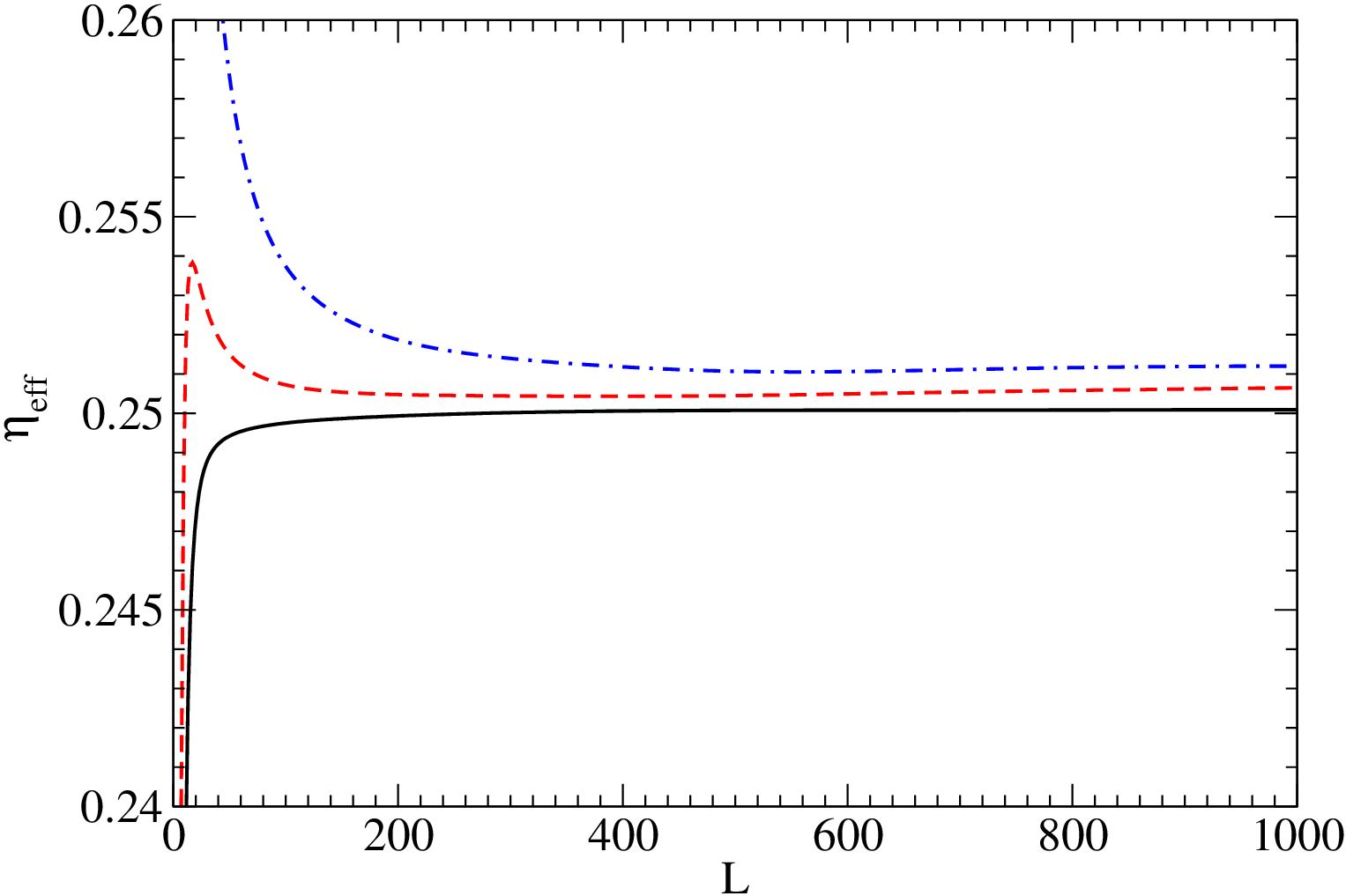}
\caption{\footnotesize (Color online) 
The WR model on the square lattice with $M=7$ components.
The dependence of $\eta_{eff}$ (\ref{eq:etaeff} ) on the system 
size $L$.
The solid, dashed and dash-dotted curves stand for 
the $(j=0, z_c=1.66196)$, $(j=0.018, z_c^L=2.58545)$ and 
$(j=0.018, z_c^U=4.47455)$ critical points, respectively.
All three curves converge to the Ising value $\eta=1/4$.
$D=300$.}
\label{fig:octo3}
\end{figure}

To confirm the Ising universality class, we have determined the critical
index $\eta$ by studying the convergence of $\eta_{eff}$, given by formula 
(\ref{eq:etaeff}), as $L\to\infty$.
The results for the WR model with $M=7$ components at the three considered 
critical points are presented in Fig. \ref{fig:octo3}.
As before, the solid, dashed and dash-dotted curves stands for 
the $(j=0, z_c=1.66196)$, $(j=0.018, z_c^L=2.58545)$ and 
$(j=0.018, z_c^U=4.47455)$ critical points, respectively.
As $L\to\infty$, all three curves converge close to the 2D Ising value 
$\eta=1/4$. 

\begin{figure}
\includegraphics[width=0.4\textwidth,clip]{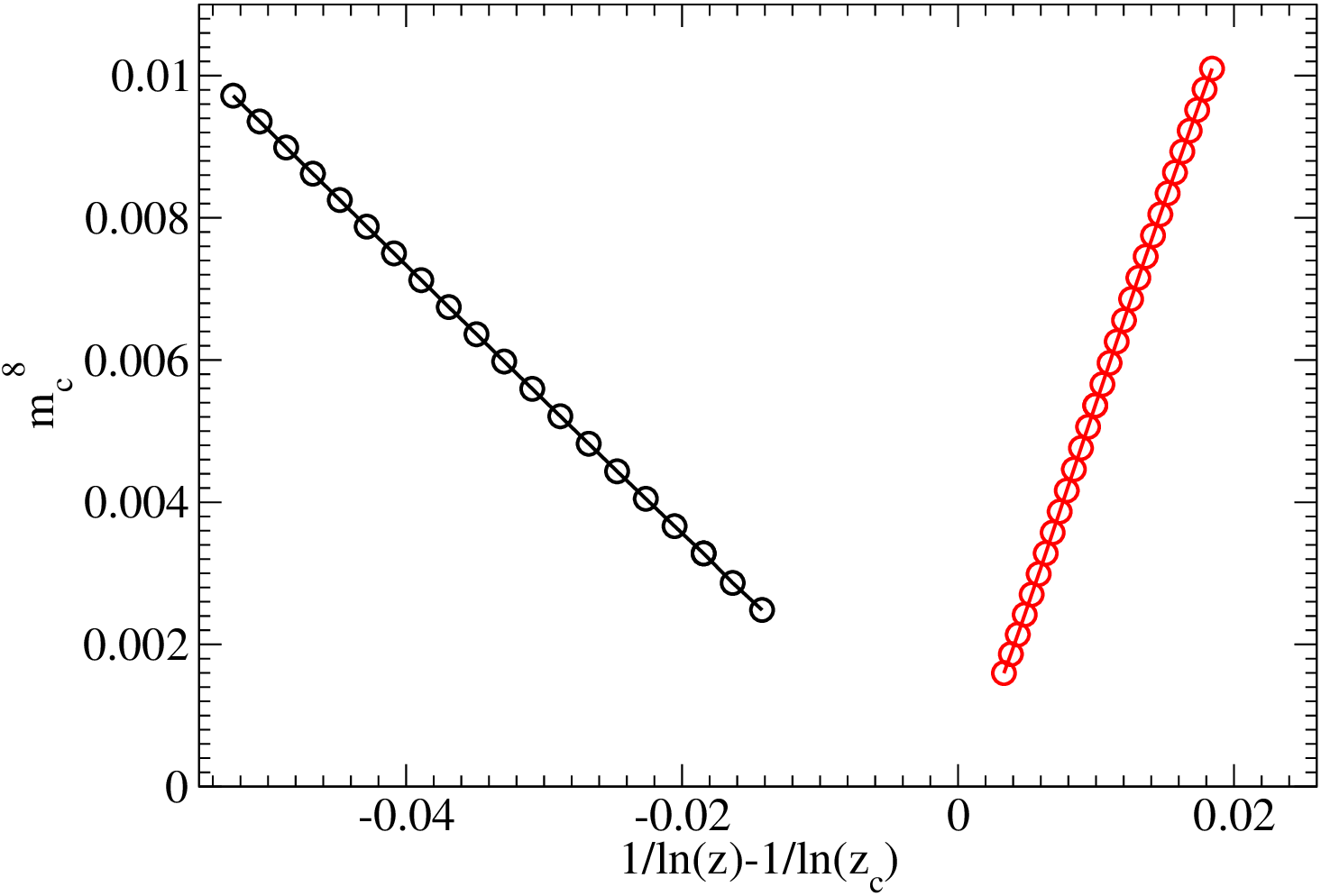}
\caption{\footnotesize (Color online) 
The generalized WR model with $M=7$ components and $j=0.018$.
The plot of $m_c^8$ as the function of $1/\ln z - 1/\ln z_c$ at
the critical points $z_c^U=4.47455$ (left data) and $z_c^L=2.58545$
(right data). 
The linear dependence confirms the Ising critical exponent $\beta = 1/8$. 
$D = 70$.}
\label{fig:octo5}
\end{figure}

As concerns the crystal order parameter $m_c$, we anticipate
the Ising behavior
\begin{equation} \label{Ising}
m_c \propto \vert T-T_c \vert^{\beta} ,\qquad  \beta=\frac{1}{8} ,
\end{equation}
close to the critical temperature $T_c$.
Since by definition the fugacity $z=\exp(\mu/k_{\rm B}T)$ with $\mu$ being
the chemical potential, it holds that $T \propto 1/\ln(z)$.
We can rewrite the relation (\ref{Ising}) as follows 
\begin{equation} 
m_c^8 \propto \vert 1/\ln z - 1/\ln z_c \vert .
\end{equation}
For the generalized WR model with $M=7$ components and $j=0.018$,
the plot of $m_c^8$ as the function of $1/\ln z - 1/\ln z_c$ is drawn in
Fig. \ref{fig:octo5}; the left data set corresponds to the critical point
$z_c^U=4.47455$ and the right data set corresponds to the critical point
$z_c^L=2.58545$.
We see that in both cases the plot is linear which confirms the Ising-like
behavior (\ref{Ising}) with the critical index $\beta=1/8$.
The knowledge of two critical exponents $\eta$ and $\beta$ determines
all remaining critical indices via the scaling relations 
\cite{Baxter,Ma,Samaj}.
 
\section{Conclusion}
The lattice version of the multicomponent WR model acquired a great deal 
of interest because it exhibits, besides the usual demixed phase transition, 
also the crystal phase driven by entropy.
The initial extremely large rigorous estimates of the minimum (number of
components to have the crystal) $M_0$ were later replaced by much smaller
values, 4 for the Bethe lattice with coordination $q=4$ and 7 for the 2D
square lattice \cite{Lebowitz95}, making the WR model of practical interest.

In the original WR model, the nearest-neighbor interaction of unlike species
corresponds to an infinite hard-core potential, with the corresponding 
Boltzmann factor $j=0$.
In real physical systems with a a finite repulsion potential, $j$ is always 
positive.
In this paper, we generalized the multicomponent WR model to an arbitrary 
repulsion $j\in [0,1]$.
The model was studied in the region close to $j=0$, namely by using 
its exact solution on the Bethe lattice (with the coordination number $q=4$) 
and the numerical CTMRG technique for the square lattice.
The consideration of nonzero $j$ has two fundamental effects on the phase
diagram.

For $M<M_0$, there is a direct transition between the disordered and demixed
phases.
In the original WR model with $j=0$, the transition is of second order 
for $M=2,3,4$. 
It was shown here that for $M=4$ the consideration of a very small value of
$j=0.001$ changes this transition to the first-order one, 
see Figs. \ref{fig:entropy} and \ref{fig:magnetization}.
For $M=3$, the second order of the disorder-demixed transition remains
unchanged for small $j$.  

The second fundamental effect concerns the crystal phase.
In the original WR model with $j=0$, if $M\ge M_0$ there are two phase
transitions as the fugacity $z$ goes from 0 to $\infty$: 
the disorder-crystal one at $z_c$ and the crystal-demixed one at $z_d$.
In the generalized WR model with $M\ge M_0$, there exists certain interval
of $j$-values in which the system undergoes three phase transitions:
the disorder-crystal one at $z_c^L$, the crystal-disorder one at $z_c^U$ 
and finally the disorder-demixed one at $z_d$. 
This reentrant disorder-disorder phenomenon is interesting not only from
an academic point of view.
If the repulsion between unlike species is too weak, the crystal phase
disappears from the phase diagram.
For all studied critical points, the generalized WR model belongs 
to the Ising universality class with the central charge $c=1/2$ and
the critical indices $\beta=1/8$ and $\eta=1/4$.

\begin{acknowledgments}
The support received from the grant QIMABOS APVV-0808-12 and the VEGA grants 
Nos. 2/0130/15 and 2/0015/15 is acknowledged. 
\end{acknowledgments}

\end{document}